\documentclass[aps,prl,twocolumn,showpacs,superscriptaddress]{revtex4-1}
\usepackage{amsmath,amssymb}
\usepackage{graphics,graphicx,color}
\usepackage{dcolumn,bm}
\usepackage{tikz}
\usepackage[colorlinks=true,citecolor=blue,urlcolor=blue]{hyperref}
\usepackage[mode=buildnew]{standalone}
\usepackage{tabularx}

\usepackage{natbib}
\usepackage{soul}
\def\[{\left[}
\def\]{\right]}
\def\({\left(}
\def\){\right)}
\def\be{\begin{equation}}
\def\ee{\end{equation}}
\def\bea{\begin{eqnarray}}
\def\eea{\end{eqnarray}}

\usepackage{soul}

\newcommand{\gaug}
{\affiliation{Institute for Theoretical Physics, Georg-August-Universit\"at G\"ottingen, 37077 G\"ottingen, Germany}}

\begin{document}
\title{Robust Prediction of Force Chains in Jammed Solids using Graph Neural Networks}

\author{Rituparno Mandal}%
\email[Email: ]{rituparno.mandal@uni-goettingen.de}
\gaug

\author{Corneel Casert}%
\email[Email: ]{corneel.casert@ugent.be }
\affiliation{Department of Physics and Astronomy, Ghent University, 9000 Ghent, Belgium}

\author{Peter Sollich}%
\email[Email: ]{peter.sollich@uni-goettingen.de}
\gaug
\affiliation{Department of Mathematics, King's College London, London WC2R 2LS, UK}

\begin{abstract}

Force chains, which are quasi-linear self-organised structures carrying large stresses, are ubiquitous in jammed amorphous materials, such as granular materials, foams, emulsions or even assemblies of cells. Predicting where they will form upon mechanical deformation is crucial in order to describe the physical properties of such materials, but remains an open question. Here we demonstrate that graph neural networks (GNN) can accurately infer the location of these force chains in frictionless materials from the local structure {\em prior} to deformation, {\em without} receiving any information about the inter-particle forces. Once trained on a prototypical system, the GNN prediction accuracy proves to be robust to changes in packing fraction, mixture composition, amount of deformation, and the form of the interaction potential. The GNN is also scalable, as it can make predictions for systems much larger than those it was trained on. Our results and methodology will be of interest for experimental realizations of granular matter and jammed disordered systems, {\it e.g.}\ in cases where direct visualisation of force chains is not possible or contact forces cannot be measured.

\end{abstract}

\maketitle

\section{Introduction}

Force chains are emergent  filament--like structures that carry large stresses when a granular material~\cite{Liu513, majmudar2005,Brodu2015, behringer2018, PhysRevLett.124.198002}, emulsion~\cite{Brujic2003,desmond2013}, foam~\cite{Katgert_2010}, dense active matter~\cite{mandal20} or assembly of cells~\cite{Delarue2016} is deformed. Unlike homogeneous simple solids, stress in such fragile matter propagates inhomogeneously via these force chains~\cite{cates1998,ostojic2006}, which therefore act as a crucial component in describing the mechanical and transport properties of such systems~\cite{PhysRevLett.84.4160, PhysRevLett.89.205501, PhysRevLett.80.61, Vandewalle_2001, Owens_2011, Smart_2007, PhysRevLett.116.188301, C5SM02326B}. Understanding when force chains will form, how the network that they make up carries the external load and responds to external or internal mechanical deformation ~\cite{geng2001,behringer05, krishnaraj2021}, and characterizing the statistical properties of force chains~\cite{nagel98, kondic2012, B926592A, PhysRevLett.100.238001, PhysRevLett.92.054302} constitute central challenges in ongoing research in granular matter systems. The study of force chains, initially qualitatively~\cite{doi:10.1143/JPSJ.5.383, DRESCHER1972337, Liu513} and later quantitatively \cite{PhysRevLett.82.5241, behringer05,behringer2018,PhysRevLett.116.078001,PhysRevLett.120.208004, Fischer2021}, became popular with the introduction of photo--elastic beads in granular matter experiments. For example, the visualisation of force chains and subsequent analysis have enabled the validation and verification of  theoretical models of granular media~\cite{Liu513}, and have helped to disentangle the distinguishing features of force chains appearing under different boundary conditions such as shear or uniform compression~\cite{majmudar2005}. 

\begin{figure*}
\includegraphics[width= .9\textwidth]{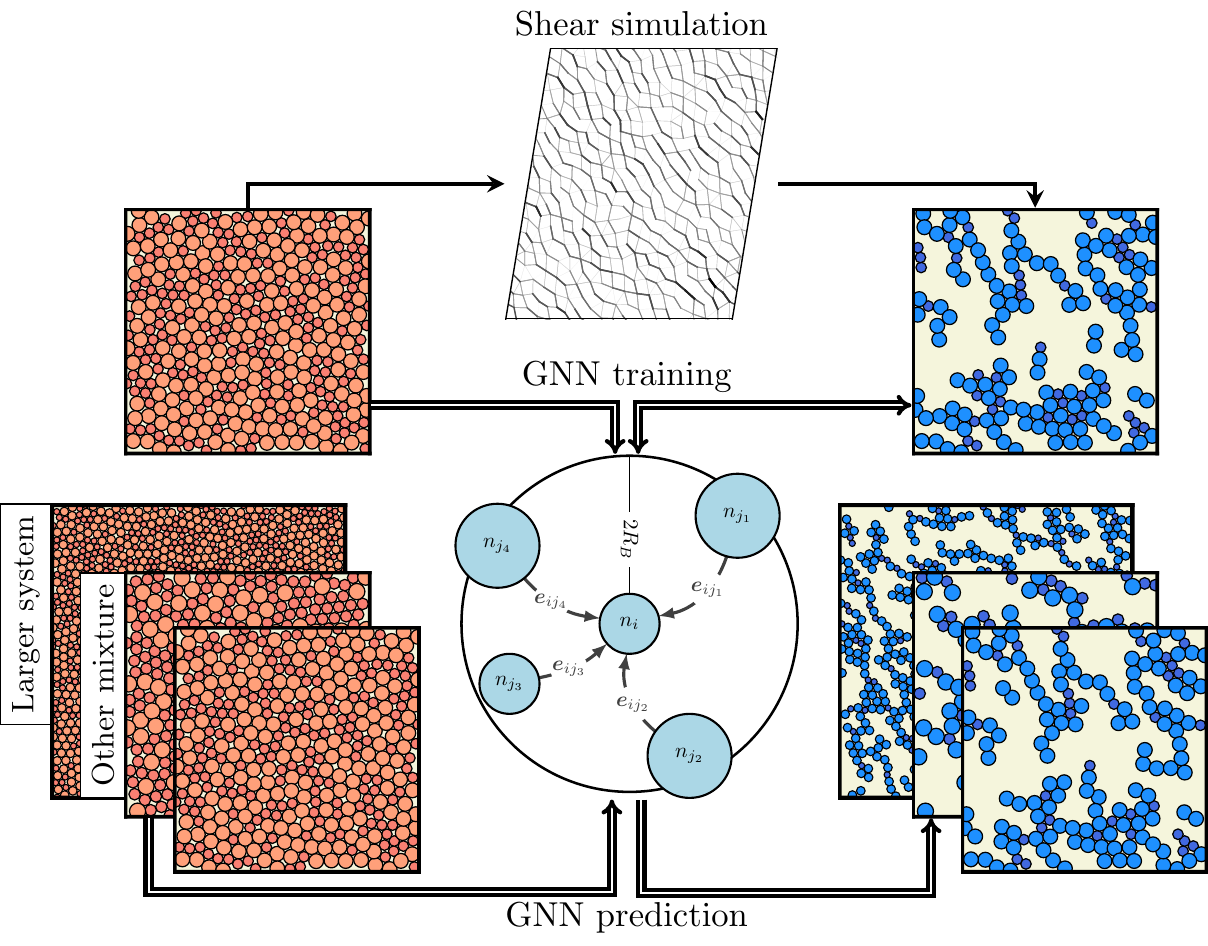}

\caption{Schematic of our method. We first generate data on the formation of force chains via a traditional method, {\textit{i.e.}} by shearing a model athermal solid in a simulation setup. We then train a graph neural network to predict the location of force chains in the deformed samples from the initial (undeformed) static structures. The trained graph neural network can then be used to predict the formation of force chains for other initial structures---even when {\em e.g.} the system size or particle mixture composition are very different from those used during training. }
\label{fig:schem}
\end{figure*}

Predicting where a force chain will arise given a deformation, \textit{i.e.}\ predicting which grains will be part of this emergent structure, is a complex problem if the interactions between the grains are unknown -- but tackling it is of vital importance in \textit{e.g.}\  material design as the force chains will be a key determinant of a material's properties~\cite{PhysRevLett.84.4160, PhysRevLett.89.205501, PhysRevLett.80.61, Vandewalle_2001, Owens_2011, Smart_2007,  PhysRevLett.116.188301, C5SM02326B}. While experimental set-ups using the aforementioned photo-elastic beads can visualise force chains~\cite{majmudar2005}, it remains impossible in numerous other experiments on granular matter, emulsions, foams etc.\ to say where force chains will form without precise knowledge of the interaction between the particles. In this article we demonstrate an efficient and accurate solution to this open question, by deploying graph neural networks (GNN) to predict the formation of force chains. 

Machine-learning methods have recently shown great potential in the analysis of physical systems, with applications ranging from quantum chemistry to cosmology \cite{carleo2019machine}. In the field of granular matter, softness was introduced as a structural predictor of regions susceptible to rearrangement, based on a classification of human-defined structure functions with a support vector machine~\cite{cubuk2015identifying, schoenholz2016structural,rocks21}, and neural networks have helped in uncovering the critical behavior of a Gardner transition in hard-sphere glasses~\cite{Lie2017392118}. 
Neural-network-based variational methods have recently also been used to study the large deviations of kinetically constrained models, which are lattice-based systems displaying glassy dynamics~\cite{whitelam2020evolutionary, casert2020dynamical}.
Graph neural networks, which operate on the elements of arbitrary graphs and their respective connectivity, have proven successful in predicting quantum-mechanical molecular properties~\cite{gilmer2017neural}, describing the dynamics of complex physical materials~\cite{sanchez2020learning}, or providing structural predictors for the long-term dynamics of glassy systems without the need for human-defined features~\cite{bapst2020unveiling}.

In this article, we show how a GNN can be trained in a supervised approach to predict the position of force chains that arise when deforming a granular system, given an undeformed static structure 
(see Fig.~\ref{fig:schem} for a schematic).
For this we first deform the system using shear deformation (step strain) and identify the particles that become part of force chains using standard methodology (see methods section). We optimize the GNN on a set of such configurations, training it to predict where the force chains will appear given the initial configurations.
We then demonstrate that the trained GNN can generalize remarkably well, allowing it to predict force chains in new undeformed samples. 
The method is extremely robust: it works exceptionally well in many different scenarios for which the GNN was not explicitly trained, which involve changes in the system size, composition, step strain amplitude, packing fraction and even interaction potential --- all without requiring any further training. 
Overall, our method provides accurate and robust predictions of where force chains will appear, without knowledge of inter-grain forces. This method can be potentially applied to numerous experiments on jammed solids, in order to determine the location of force chains even when it is not possible to visualize them directly.  

\section{Method}
\label{method}

\subsection{Model \& Simulation}

We train our GNN on configurations of a frictionless~\cite{PhysRevE.62.2510, PhysRevE.60.687, Head2001, PhysRevLett.97.258001, PhysRevE.96.042901} athermal soft binary sphere mixture with harmonic interactions~\cite{DurianPRL1995,DurianPRE1997,Chacko2019} as a model athermal solid. 
The pair-interaction potential between two particles $i$ and $j$ is completely repulsive, and of the form 
\begin{equation}\label{eq:pothar}
    V(r_{ij}) =  \frac{1}{2} G R^3 \left(1-\frac{r_{ij}}{R_i+R_j}\right)^2
\end{equation}
when particles overlap and zero otherwise. Here $R_i$ and $R_j$ are the radii of the $i$-th and $j$-th particle respectively and $R$ is a characteristic particle radius parameter. We set $G=1$, $R=1$ throughout all simulations. Our initial simulations (for training data generation) are performed using a binary mixture of $N = n_A + n_B$ particles with $n_A=n_B=200$ and $R_A=R$, $R_B=1.4 R$ in a two-dimensional box of linear size $L$ and with periodic boundary conditions.
For these training data generation runs, the simulations are performed at a fixed packing fraction $\phi$, which is defined as $\phi=(n_A \pi R_{A}^2+n_B \pi R_{B}^2)/L^2$. 
We always start with a force-free configuration under athermal conditions, and then deform the system by applying a step shear strain of magnitude $\gamma$. Lees-Edwards periodic boundary conditions are used to implement the step strain. 
\begin{figure}
\includegraphics[width=\columnwidth]{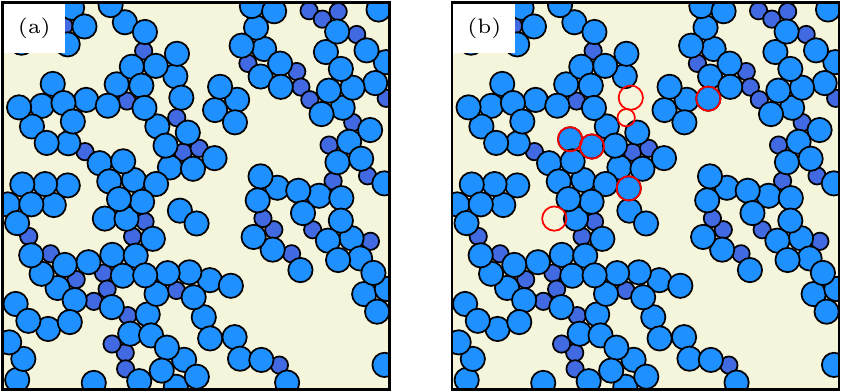} 
\caption{(a) Force chains (blue) in an athermal solid with $n_A = n_B = 200$, obtained through a shear simulation with step strain $\gamma = 0.1$ at packing fraction $\phi=1.0$. (b) Force chains predicted by a graph neural network taking as input the configuration of (a) before deformation. Particles misidentified by the GNN are highlighted in red. }
\label{fig:compare}
\end{figure}

\subsection{Identification of Force Chains}

To quantitatively detect force chains we follow the approach detailed in \cite{peters2005characterization,tordesillas2010force}, where force chains are defined as quasi-linear structures formed of particles that carry above-average load, {\em i.e.}\ compressive stress. 
To identify the particles within the force chains, we first calculate the stress tensor $\hat{\sigma}_{\alpha\beta} = \sum_{i=1}^{N_{nb}} f_{\alpha}^i r_{\beta}^i$ for each particle in the instantaneously sheared configuration, where $N_{nb}$ is the number of neighbouring particles exerting a force on the central particle, and $f_{\alpha}^i$ and $r_{\beta}^i$ are the components of the force and the 
radius vector connecting the centers of the two interacting particles, respectively. 
The largest eigenvalue of $\hat{\sigma}_{\alpha\beta}$ is the magnitude of the particle load vector, while its orientation is given by the corresponding eigenvector. 
Neighbouring particles whose load vectors align within an angle of $45^\circ$ and have above-average (arithmetic mean) magnitude 
are assigned as part of a force chain \cite{peters2005characterization,tordesillas2010force}. In a completely analogous fashion one could train neural networks on force chains identified using other approaches, \textit{e.g.}\ based on community detection~\cite{C4SM01821D, Huang2016} or topological properties of the force network~\cite{PhysRevE.87.042207}.
Note that these detection techniques~\cite{peters2005characterization,tordesillas2010force, C4SM01821D, Huang2016, PhysRevE.87.042207}  allow for the analysis of the force network, but do not provide any predictive power on how it will change upon deforming a system. 
\begin{figure}
\includegraphics[width=\columnwidth]{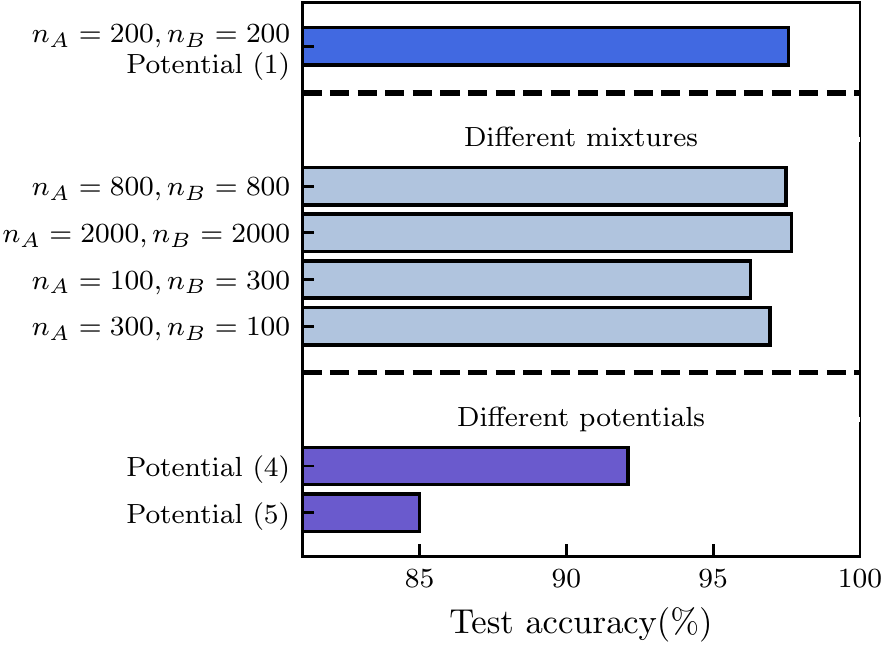} 
\caption{Robustness of force chain prediction accuracy: a graph neural network trained on a data set with a specific potential and number of particles (top) can also accurately predict force chains for configurations with a different system size (number of particles), different ratio of mixture components or with a different interaction potential. All results shown here are obtained at packing fraction $\phi = 1.0$.}
\label{fig:acc}
\end{figure}

\subsection{Training of GNN and prediction of Force Chains}

To predict the location of force chains after a deformation with graph neural networks, we first transform a given initial configuration to a graph by drawing edges between particles that are within a fixed cutoff-distance (here set to $2R_B$).
Each node of this graph has the corresponding particle radius as feature $n_0$. 
When conditioning our graph neural network on a global property such as the magnitude of the deformation $\gamma$ or packing fraction $\phi$, we include this property as an additional (uniform) node feature.
The features assigned to each edge of the graph $e_{ij}$ consist of the distance between the two particles connected by the edge and the unit vector in the direction of their relative displacement. Importantly, we do not need to include any knowledge of the contact forces between the particles, which are typically difficult to measure~\cite{PhysRevLett.116.078001, PhysRevLett.117.159801}.
\begin{figure*}
\includegraphics[width=.9 \textwidth]{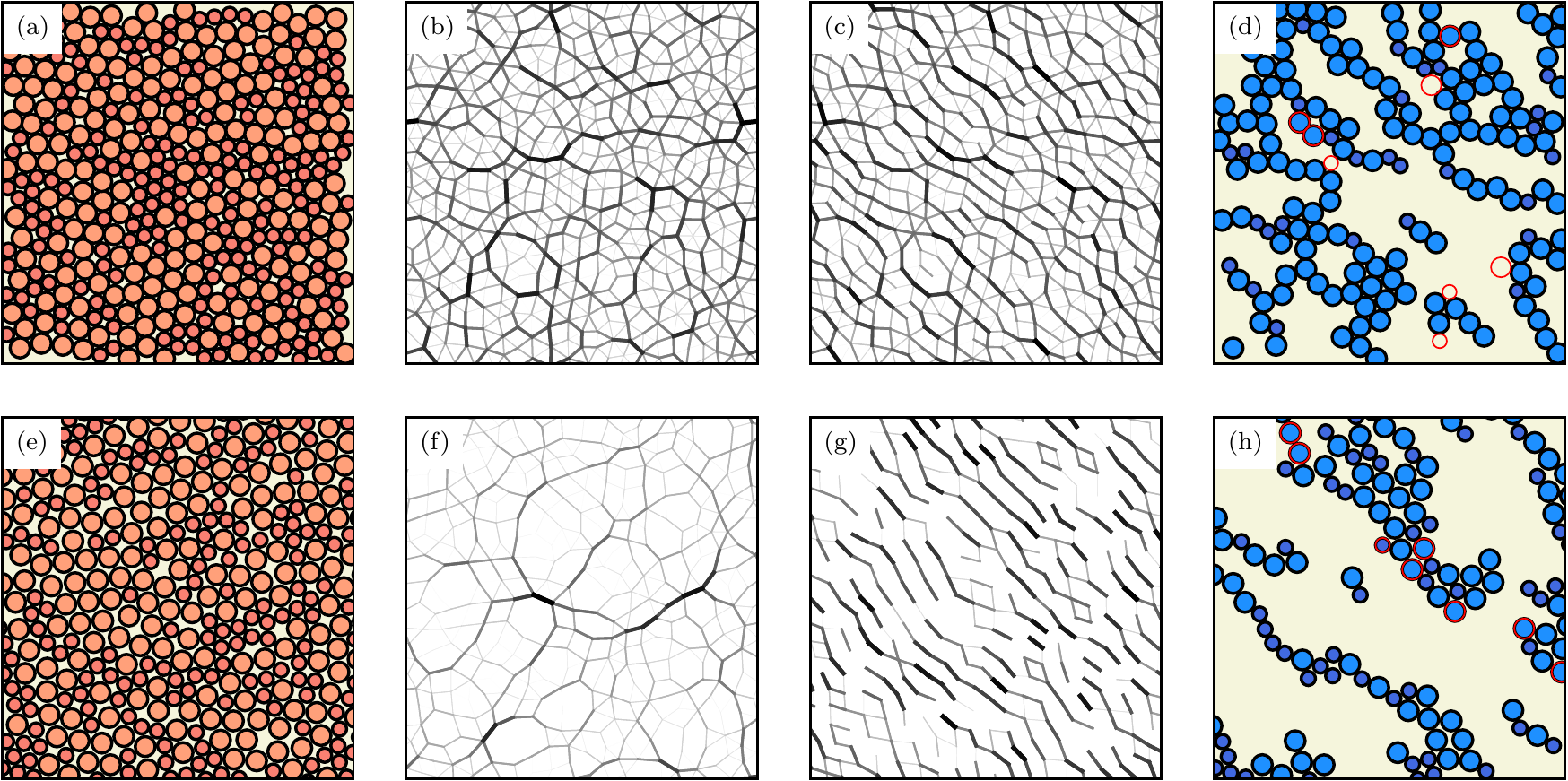} 
\caption{ (a) Example configuration before deformation with $n_A = n_B = 200$ particles interacting through a harmonic potential (as given in Eq.~(\ref{eq:pothar})) at packing fraction $\phi = 1.0$. (b) Corresponding force network. (c) Force network after a deformation with step strain of amplitude $\gamma = 0.1$. (d) Force chains of the configuration in (c). Differences between simulation and GNN prediction on the basis of (a) are highlighted in red. (e-h) Same as in (a-d) but now for a configuration at a packing fraction ($\phi = 0.845$) close to jamming.  }
\label{fig:networks}
\end{figure*}\\
In order to predict which particles will become part of force chains after deforming the initial configuration, we apply a graph neural network to this graph.
Such a GNN consists of $N_l$ layers, where in every layer $l$ the node features corresponding to each particle $i$ are updated according to the features of the particles in their neighbourhood $\mathcal{N}(i)$ and the features of the edges that connect them (see schematic in Fig.~\ref{fig:schem}).
Specifically, 
\begin{equation}
\label{eq:cgcnn}
(n_l)_i = (n_{l-1})_i + \sum_{j \in \mathcal{N}(i)} f^l_\mathcal{W}\left[(n_{l-1})_i,(n_{l-1})_j, e_{ij}\right],
\end{equation}
where $f^l_\mathcal{W}$ is a parameterized non-linear function (neural network) that calculates new node features for each particle. Importantly, we use the same function $f^l_\mathcal{W}$ for each particle, which allows us to apply the GNN to systems with an arbitrary number of particles. 

For our force chain prediction, we pass the features $n_{N_l}$ of each node in the final layer through a fully-connected neural network with one output feature and a sigmoid activation function. This final result is, for each particle, the probability $\hat{p}$ of being part of a force chain after deformation, as assigned by the neural network.

During training we optimize the weights of the GNN such as to minimize the cross-entropy \begin{equation}
    \mathcal{H}(p,\hat{p}) = -\frac{1}{N}\sum_{i=1}^N p_i \log(\hat{p}_i) + (1-p_i)\log(1-\hat{p}_i),
\end{equation}
where $p=1$ if a particle is part of a force chain in a particular training configuration, and $p=0$ otherwise.
We use the Adam optimizer~\cite{kingma2014adam} to minimize this loss function on a training set, and choose the model hyperparameters as those that perform best on a validation set consisting of configurations not included in the training set. We then evaluate our network on an independent test set. The optimized hyperparameter values and more details on the neural network architecture and training can be found in the SI.

\section{Results}

We first demonstrate that an optimized GNN can predict very accurately where force chains will form for the canonical case of a jammed solid consisting of a binary mixture of harmonic particles (as described in Methods) at $\phi=1.0$. 

In Fig.~\ref{fig:compare}, we compare the force chains obtained through a numerical shear simulation (Fig.~\ref{fig:compare} (a)) to those predicted by our trained GNN (Fig.~\ref{fig:compare} (b)), which receives an undeformed configuration as input.
Only a few particles (highlighted in red) are misidentified by the GNN,  indicating its very high prediction accuracy, as described in Fig.~\ref{fig:acc}.
In the following sections, we describe the most useful aspects of the GNN predictor for the prediction of force chains, namely scalability and robustness.

\subsection{Scalability}

As each graph-convolutional operation in our GNN only depends on a particle's local neighbourhood (Eq.~\ref{eq:cgcnn}), the GNN is not explicitly dependent on the number of nodes in the graph.
This allows us to apply the GNN on systems containing a much larger number of particles than it was initially trained on, as long as the relevant physical length-scales stay of the same order when increasing system size.
Doing so allows for massive numerical acceleration in the study of large-scale systems, as the largest computational cost for training the GNN lie in the generation of a large enough training data set and in the optimization of the network's hyperparameters --- both of which are much more expensive when training on large systems.
We demonstrate the success of this strategy in Fig.~\ref{fig:acc}, where we train a GNN on data acquired for a small system size ($N=400$) and then apply it to much larger systems (shown up to $N=4000$) without a decrease in force chain prediction accuracy. This is important in the context of predicting force chains in experiments: once trained, evaluating our prediction method only demands a cost that increases linearly with system size so that one can easily make predictions on the much larger systems one has to deal with in a typical experiment. More results on scalability near jamming are provided in the SI.

\begin{figure}
\includegraphics[width=\columnwidth]{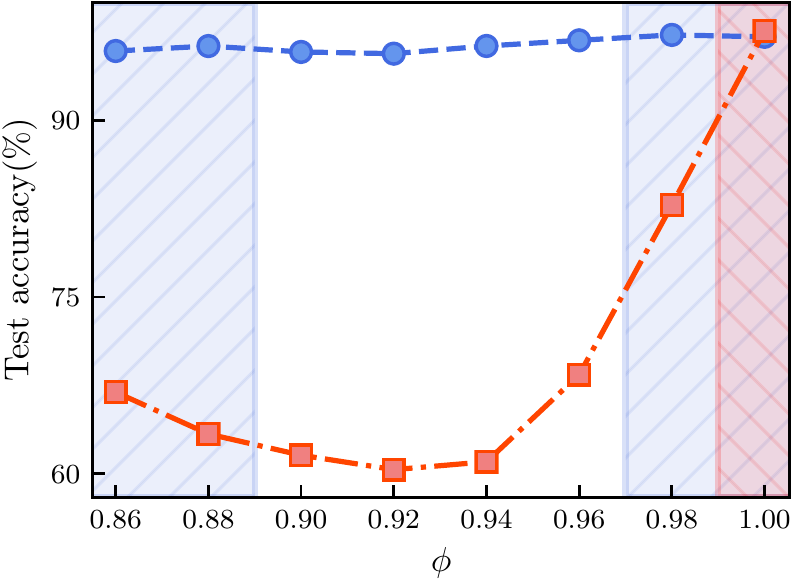} 
\caption{Force chain prediction accuracy for systems at \mbox{$\gamma = 0.10$} and with different packing fractions, ranging between $\phi=0.85$ and $\phi=1.0$, with a GNN conditioned on $\phi$. We show results obtained with a GNN trained on a data set obtained at $\phi = 1$ (red), and a GNN trained on data of samples obtained at $\phi\le0.88$ and $\phi\ge 0.98$ (blue). The latter GNN provides highly accurate predictions in the interpolation regime $0.9 \le \phi \le 0.96$, even though it was never trained on data obtained at these values of the packing fraction.}
\label{fig:int_phi}
\end{figure}

\subsection{Robustness}

We first observe  that a GNN can predict force chains very accurately when either trained on a data set generated at high ($\phi=1.0$) or low packing fractions ($\phi=0.845$, close to jamming); Figs.~\ref{fig:networks}(a) and (e) show sample configurations.
Two separate GNNs were trained on data obtained at the two packing fractions. Figs.~\ref{fig:networks}(d),(h) show their predictions (on previously unseen, undeformed configurations) of force chains, which match remarkably well with the direct simulations at both high and low packing fraction.
This consistent accuracy is achieved 
even though the force networks at the different packing fractions have a very distinct structure, as can be seen from Figs.~\ref{fig:networks} (b),(f) and (c),(g) showing the configurations before and after deformation, respectively.

\begin{figure}
\includegraphics[width=\columnwidth]{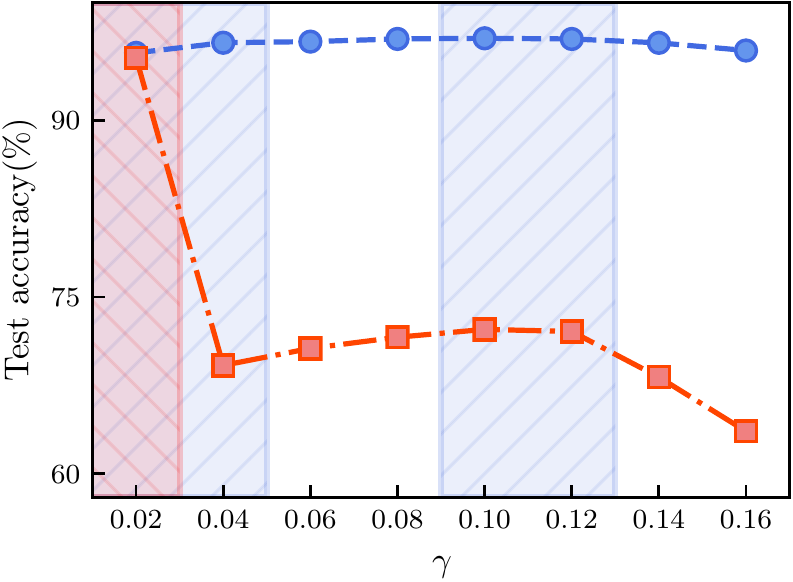} 
\caption{Force chain prediction accuracy for different step strain amplitudes $\gamma$: this is the analogue of Fig.~\ref{fig:int_phi}, now for GNNs conditioned on $\gamma$. The GNN trained on only one value of $\gamma = 0.02$ fails to predict force chains accurately for larger deformations (red), while a GNN trained on $\gamma \le 0.04$ and $0.10 \le \gamma \le 0.12$ (blue) successfully interpolates between these deformation magnitudes; it also extrapolates reliably to larger values of $\gamma$.}
\label{fig:int_gamma}
\end{figure}

A network trained on a fixed, single value of the packing fraction $\phi$ provides inaccurate predictions when applied to configurations obtained at other packing fractions (see Fig.~\ref{fig:int_phi}). If, however, we provide the GNN with information about the packing fraction during training, it can correctly predict force chain formation over a wide range of values for the packing fraction, even for values not included in the training data (see Fig.~\ref{fig:int_phi}).
Likewise, we can provide the neural network with the magnitude of the step strain $\gamma$, and apply it to correctly predict force chains when varying the value of $\gamma$ for a given configuration. Again, this works even for values of $\gamma$ for which the GNN did not observe any data during optimization: in Fig.~\ref{fig:int_gamma} we demonstrate that the GNN produces highly accurate predictions both when interpolating between values of $\gamma$ included in the training set, and when extrapolating towards higher values of $\gamma$.

The trained GNN is also robust to changes in the composition of the binary mixture. To demonstrate this, we generate samples with a very different composition by changing the ratio of $n_A:n_B$, while keeping the packing fraction fixed. This is demonstrated in Fig.~\ref{fig:acc} for the cases of $n_A:n_B=3:1$, and $n_A:n_B=1:3$, with a GNN that was originally trained on data with $n_A:n_B=1:1$. The predictions are remarkably good for both cases, with only a minimal loss of accuracy when compared to the original mixture composition.

The fact that a GNN can provide predictions without any significant loss in accuracy in these three different generalization scenarios is highly advantageous, as it implies that large amounts of data only need to be obtained (either through experiment or simulation) at a few model parameter combinations, thus greatly reducing the cost for predicting where force chains will form in a generic experiment.

Finally, we ask our GNN to predict the force chains in a jammed solid with different pairwise interactions---either through a Hertzian potential \begin{equation}\label{eq:pothertz}
    V(r_{ij}) =  \frac{2}{5} G R^3 \left(1-\frac{r_{ij}}{(R_i+R_j)}\right)^{{5}/{2}},
\end{equation} 
or a power-law potential 
\begin{equation}\label{eq:potpl}
    V(r_{ij}) = \epsilon \left(\frac{R_i+R_j}{r_{ij}}\right)^{10},
\end{equation}where we set $\epsilon=1$.
We find that even in these cases, the GNN retains most of its accuracy (Fig.~\ref{fig:acc}), with a larger performance reduction for the power-law potential. This robustness to the nature of the interaction potential is extremely important for the applicability of our method in an experimental context, where the particle interactions might not be straightforward to estimate.

\section{Discussion}    

In this article, we made use of graph neural networks (GNN) to accurately predict where force chains will arise upon deformation of jammed disordered solids. Our network is trained on data obtained through direct shear simulations, and exhibits very high generalization ability to unseen configurations. Crucially, the optimized GNN is robust to changes in a variety of model parameters, such as the packing fraction, system size, magnitude of deformation and interaction potential: it produces very accurate predictions for such cases, without having ever observed them during its training. Although we have used data obtained through numerical simulations to train our GNN, the methodology would be identical for experimental data, where it is not always possible to measure inter-particle forces. We have verified (see Supplementary Information for details) that the predictions of our GNN are not correlated with the local $D^2_{\rm min}$, an indicator of plastic rearrangements. This can in turn be predicted using softness, which is a machine learning feature that has recently been used extensively~\cite{cubuk2015identifying, schoenholz2016structural, rocks21}. Overall, this indicates that our GNN is picking up on novel features in the local structure of such disordered solids.

Our study will open new possibilities in experiments on such disordered solids (granular matter, emulsions, foams, etc.) where direct visualisation of force chains is not possible, allowing for more in-depth analysis of structural properties as quantified by force chains. Even though for the demonstration of the success of our method we have used data generated by simulations of two-dimensional frictionless grains, as is common in numerical simulations~\cite{PhysRevE.62.2510, PhysRevE.60.687, Head2001, PhysRevLett.97.258001, PhysRevE.96.042901},  it is straightforward to extend the method to scenarios where frictional interactions between the grains are present~\cite{ostojic2006}, as well as to three-dimensional configurations where visualisation of force chains in experiments is a very challenging task~\cite{Zhou1631}.

RM and CC contributed equally to this work. We are grateful to Bulbul Chakraborty for valuable comments about the manuscript. This project has received funding from the European Union’s Horizon 2020 research and innovation programme under the Marie Skłodowska-Curie grant agreement No 893128. Computational resources (Stevin Supercomputer Infrastructure) and services used in this work were provided by the VSC (Flemish Supercomputer Center), and the Flemish Government – department EWI.

\bibliography{mlforcechain}

\clearpage
\newpage

\appendix*

\section*{Supplemental Information for ``Robust Prediction of Force Chains in Jammed Solids using Graph Neural Networks''}
\section{Model architecture and training details}

Given an undeformed configuration, the first step in our force-chain prediction routine is to create a graph by identifying the particles as nodes and drawing edges between nodes that are separated by a maximum distance of $2R_B$. 
We assign a feature vector $e_0$ to each edge, which contains the distance and relative orientation of the particles it connects; and a feature vector $n_0$ to each node, which contains the particle radius and any global features being conditioned on, such as the packing fraction or the amplitude of the step strain deformation.
We then embed the node features $n_0$ and edge features $e_0$ of the initial graph in a higher-dimensional space using a parameterized linear transformation
\begin{align}
    n_0^\prime &= W^{(n)}n_0 + b_n, \\
    e_0^\prime &= W^{(e)}e_0  + b_e.
\end{align}
Here, $W^{(n)}$ and $W^{(e)}$ are weight matrices of dimension $( d_h\times |n_0|)$ and $(d_h\times|e_0| )$ respectively and $b_n$ and $b_e$ are $d_h$-dimensional bias vectors; $d_h$ is a hyperparameter called the hidden dimension.
We then pass these features through $N_l$ layers of a residual graph-convolutional network.
In each of the layers $l$, we calculate new node features $n_l^\prime$ for each particle $i$ from the features of its neighbouring particles $\mathcal{N}(i)$ and the connecting edges, by applying a linear transformation followed by a non-linear activation function.
The features $n_l^\prime$ are then added to the original features  $n_{l-1}^\prime$, which allows for more stable training.
Explicitly, we use the graph-convolutional operator~\cite{xie2018crystal}
\begin{equation}
\begin{split} \label{eq:cgcnn2}
    (n_l^\prime)_i = &(n_{l-1}^\prime)_i +\\ &\sum_{j\in\mathcal{N}(i)} \sigma (W^f_{l} (z_{l-1})_{i,j} + b_l^f) \odot g(W^s_{l} (z_{l-1})_{i,j} + b_l^s).
    \end{split}
\end{equation}
Here, $(z_{l-1})_{i,j}$
is the vector obtained by concatenating the node features $(n_{l-1}^\prime)_i$ and $(n_{l-1}^\prime)_j$ with the edge features 
$(e_0^\prime)_{i,j}$, $\sigma(x) \equiv 1/(1+e^{-x})$ and $g(x) \equiv \ln(1+e^x)$ are a sigmoid and softplus 
function respectively, and $\odot$ represents element-wise multiplication. 
$W^f_{l}$ and $W^s_{l}$ are trainable weight matrices 
of size $d_h \times 3d_h$ acting on the node and edge features, while $ b_l^f$ and  $b_l^s$ are $d_h$-dimensional bias vectors. Note that the weights and biases are identical for each node. This type of graph-convolution provided the most accurate predictions in our numerical experiments; other graph convolutions we tested resulted in a lower accuracy, including the graph-convolutions used in~\cite{li2020deepergcn, hu2019strategies, gilmer2017neural}  (implemented using~\cite{Fey/Lenssen/2019}).\\
\begin{figure}
\begin{center}
\includegraphics[width=\columnwidth]{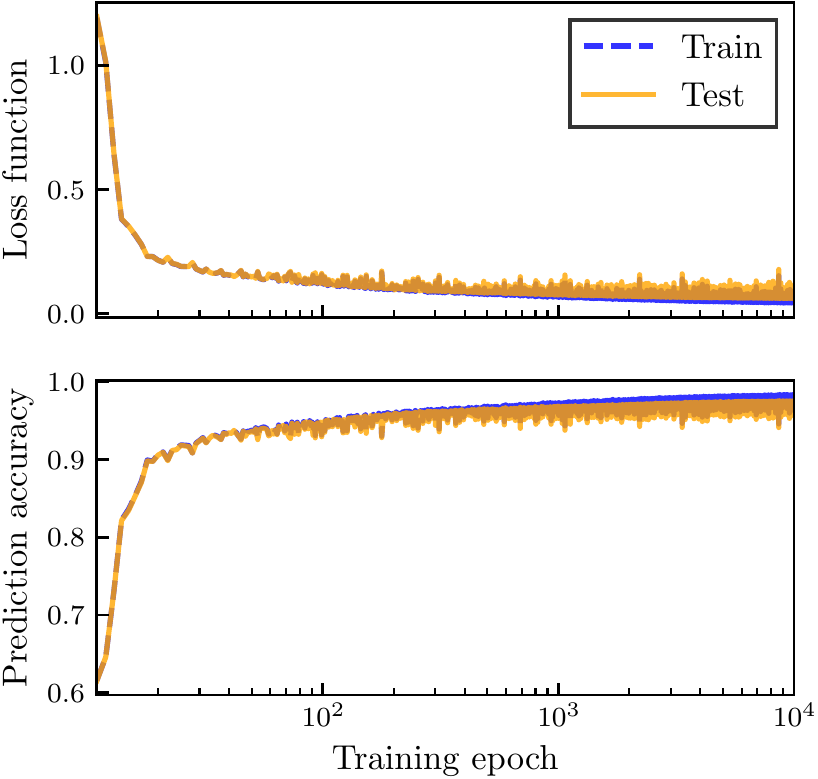}
\caption{Loss function (cross entropy) and prediction accuracy during training, evaluated on a training set consisting of 1024 configurations and a testing set consisting of 128 configurations.}
\label{fig:training}
\end{center}
\end{figure}

To optimize the weights and biases, we minimize the cross-entropy \begin{equation}
    \mathcal{H}(p,\hat{p}) = -\frac{1}{N}\sum_{i=1}^N p_i \log(\hat{p}_i) + (1-p_i)\log(1-\hat{p}_i),
\end{equation}
where $p=1$ if a particle is part of a force chain, and $p=0$ otherwise. The weight updates are performed using the Adam optimizer~\cite{kingma2014adam}. In Fig.~\ref{fig:training}, we show how the loss decays and the accuracy increases (both on a training and testing set) during a typical training routine.   \\
\begin{figure}[t]
\includegraphics[width=\columnwidth]{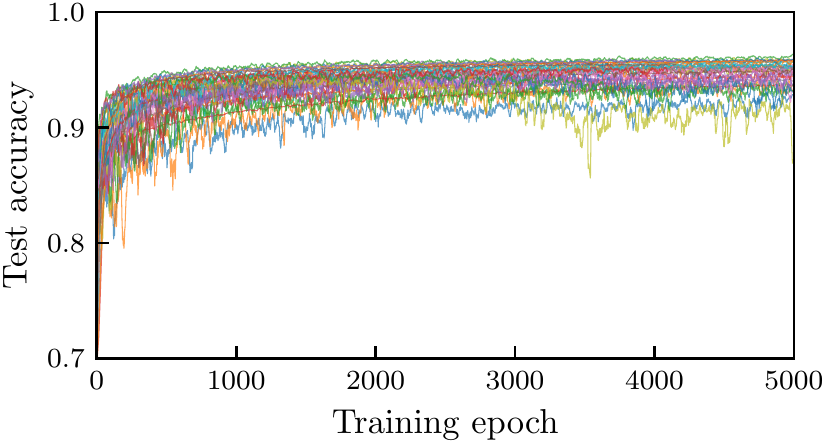} 
\caption{Evolution of the prediction accuracy on a validation set during training of the GNN. Each line represents a different combination of hyperparameter values. We considered values of $d_h$ between 16 and  128, of $N_l$ between 2 and 12, and of $\alpha$ between $0.0001$  and $0.1$.}
\label{fig:hyperparams}
\end{figure}
\begin{figure}
\begin{center}
\includegraphics[width=\columnwidth]{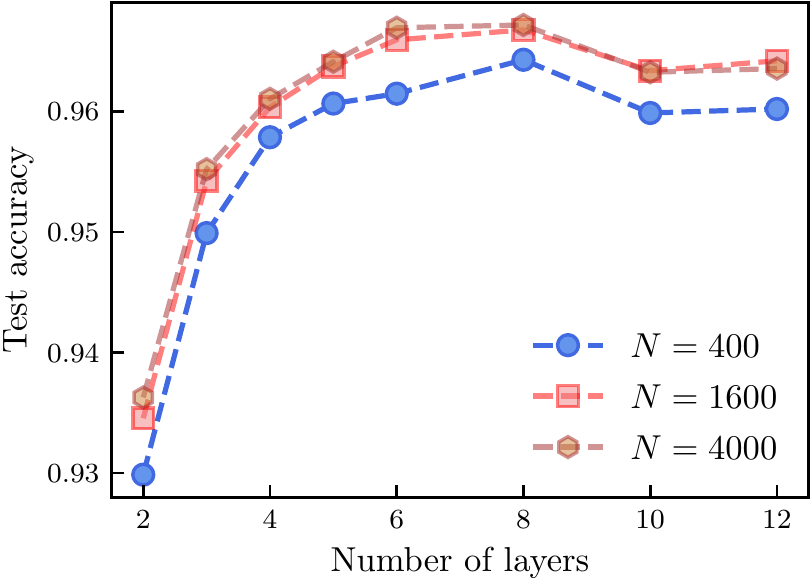}
\caption{Test accuracy for systems at $\phi=0.845$ with number of particles between 400 and 4000, as a function of the number of layers of the GNN.}
\label{fig:nlayers}
\end{center}
\end{figure}

In order to choose the model and training hyperparameters (\textit{i.e.}\ the learning rate $\alpha$ for the Adam optimizer, the hidden dimension $d_h$ and the number of layers $N_l$; we used a fixed batch size of 64 configurations) that provide the most accurate predictions, we optimized a GNN for many combinations of these hyperparameters on a data set of 1024 configurations of $n_A = n_B = 200$ 
particles at a packing fraction of $\phi = 1.0$. We then evaluated the loss function on an independent validation set consisting of 128 configurations, and chose the hyperparameters that resulted in the lowest loss: $d_h =64,~N_l = 8,~\alpha = 0.001$. All results shown in the manuscript were obtained with this set of hyperparameters. We used 1024 training configurations for each set of different values of the control parameters (\textit{i.e.}\ combination of packing fraction $\phi$ and amplitude of step strain $\gamma$), while the reported test accuracy is evaluated on 128 independent configurations.

We note here that our method is quite robust against changing the values of the hyperparameters mentioned above, and GNNs containing far fewer hyperparameters did not result in significantly worse prediction accuracy. In Fig.~\ref{fig:hyperparams}, we visualize how the accuracy on the validation set increases during training for many different hyperparameter combinations; each tested combination resulted in an accuracy of at least $90\%$.\\

One concern was that the influence of long-range correlations near jamming (\textit{i.e.} $\phi=0.845$) might negatively affect our ability to obtain accurate predictions for systems larger than those trained on, or require excessive network depth to incorporate these long-range correlations. In Fig.~\ref{fig:nlayers}, we demonstrate that this is not the case: even very shallow GNNs consisting of just two layers can predict force chains in systems containing ten times as many particles as seen during training, with an accuracy of $93\%$.\\

\section{Comparison between the force chain locations and $D^2_{\text{min}}$}

Previous machine learning studies on glassy systems have mainly focused on the calculation of `softness', a structural predictor of regions susceptible to rearrangement~\cite{cubuk2015identifying, schoenholz2016structural, rocks21}. It has been demonstrated quite extensively that softness is strongly correlated with $D^2_{\text{min}}$~\cite{PhysRevE.57.7192}, which quantifies the magnitude of nonaffine displacement during a time interval $\Delta t$. 
For each particle $i$, it is calculated as \begin{equation}
    D^2_{\text{min}}(i) = \min_\Lambda \frac{1}{|\mathcal{N}(i)|}\sum_{j \in \mathcal{N}(i)} (\mathbf{r}_{ij}(t+\Delta t) - \Lambda \mathbf{r}_{ij}(t))^2,
    \label{eq:d2min}
\end{equation}
where the sum runs over the neighbours $\mathcal{N}(i)$ of particle $i$,  $\mathbf{r}_{ij}$ is the displacement between particles $i$ and $j$, and we minimize over the possible local strain tensors $\Lambda$.
We also consider an instantaneous version of $D^2_{\text{min}}$~\cite{Chacko2019}\begin{equation}
    D^{2'}_{\text{min}}(i) =  \min_\Theta \sum_{j \in \mathcal{N}(i)}[(\mathbf{v}_{i}-\mathbf{v}_{j}) - \Theta (\mathbf{r}_{i}-\mathbf{r}_{j})]^2.
    \label{eq:d2min_inst}
\end{equation}

\begin{figure}[b]
\begin{center}
\includegraphics[width=\columnwidth]{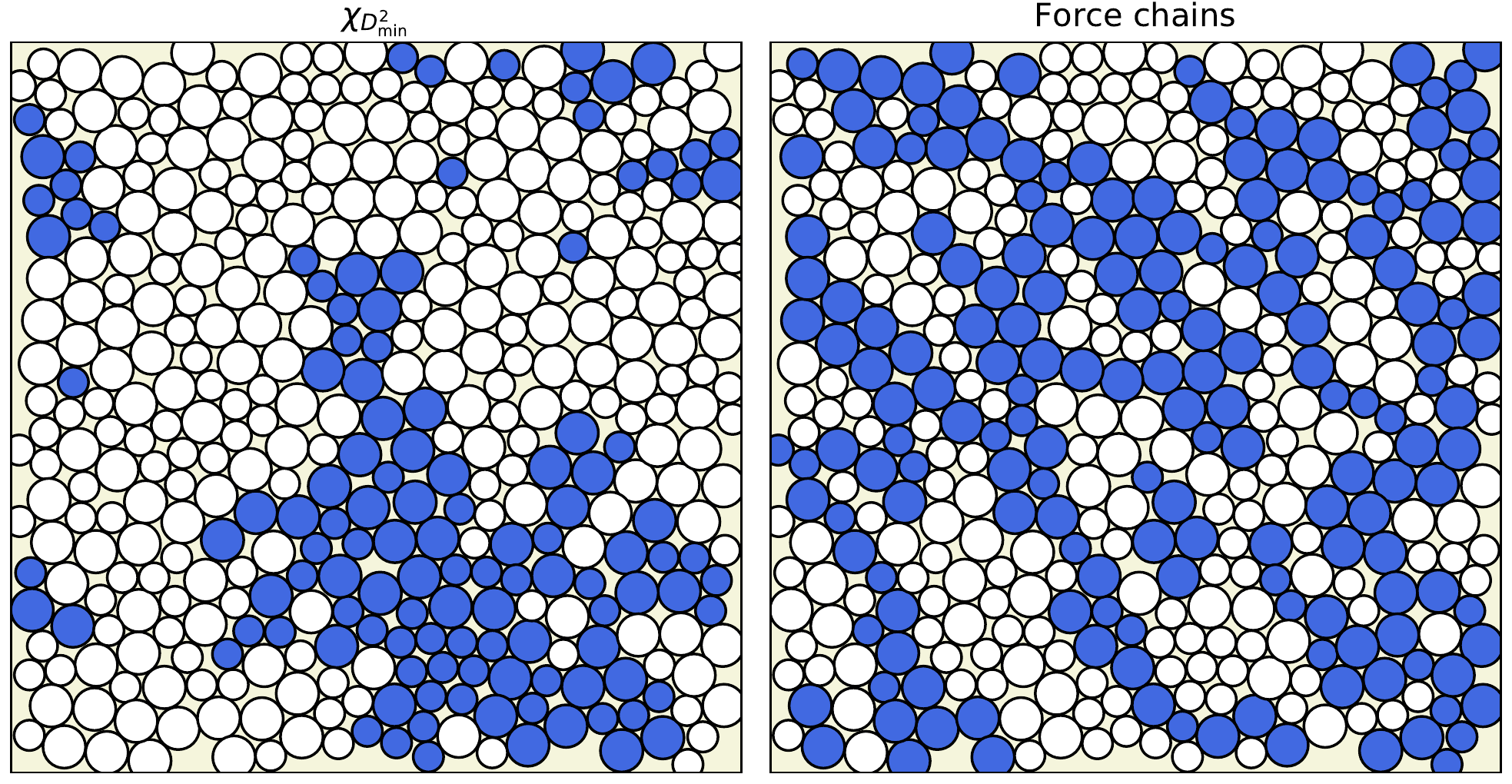}
\caption{Left: thresholded version $\chi_{D^2_\text{min}}$ of $D^2_\text{min}$ (Eq.~(\ref{eq:d2min})): particles that had values of $D^2_\text{min}$ in the top 30\% before deformation are shown in blue. Right: Particles classified as being part of a force chain after deformation are shown in blue.}
\label{fig:d2min}
\end{center}
\end{figure}

\begin{figure}[b]
\begin{center}
\includegraphics[width=\columnwidth]{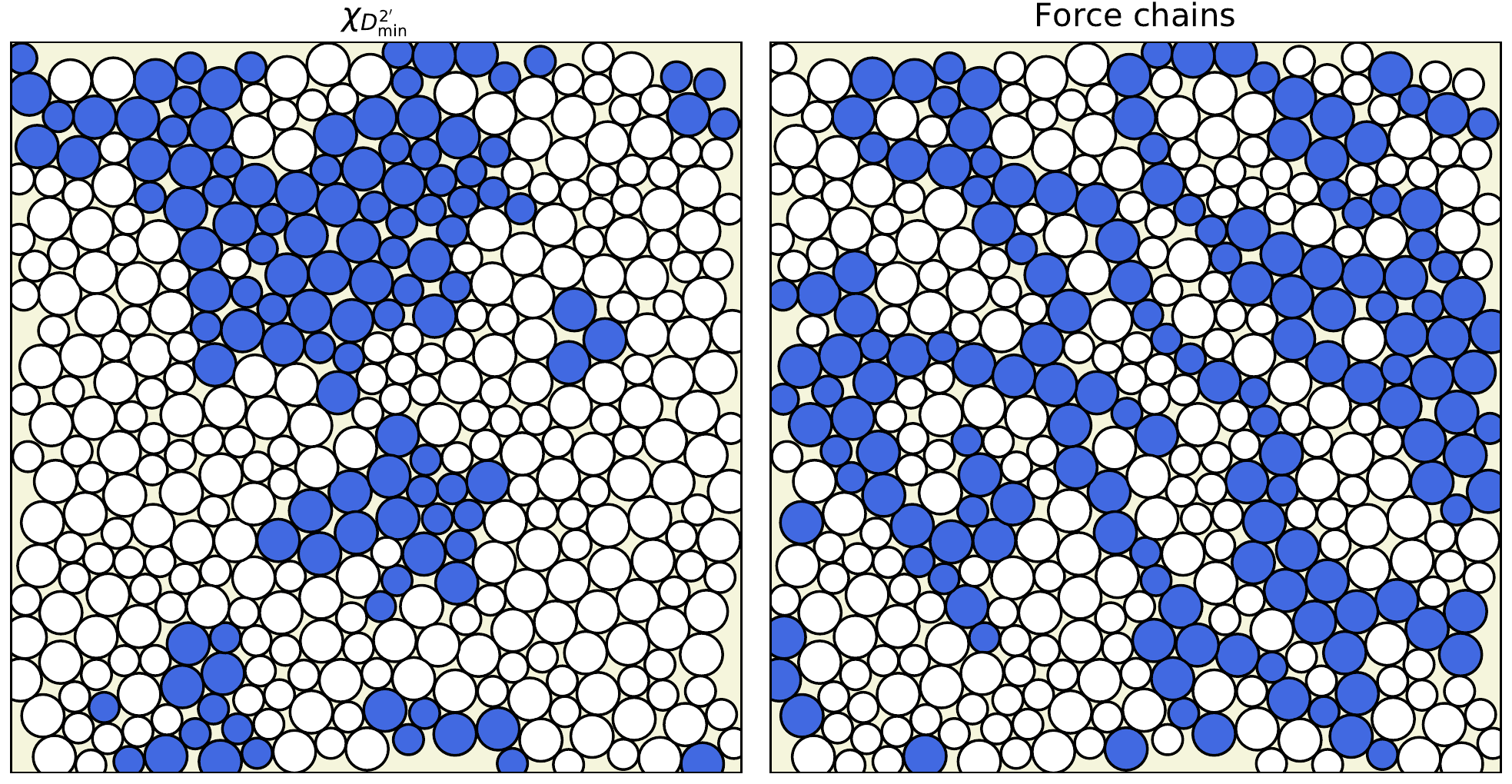}
\caption{Same as Fig.~\ref{fig:d2min}, but with the instantaneous version of $D^2_\text{min}$, as in Eq.~(\ref{eq:d2min_inst}).}
\label{fig:d2min_inst}
\end{center}
\end{figure}

In Fig.~\ref{fig:d2min} and Fig.~\ref{fig:d2min_inst}, we demonstrate that our force chain prediction is not correlated to the local $D^2_{\text{min}}$ (for both definitions) in the configuration prior to deformation. This indicates that our GNN has identified novel features in the structure of these configurations, which were not captured in earlier machine-learning based studies.


\end{document}